\documentclass[10pt]{iopart}
\usepackage{iopams}  
\usepackage{graphicx}
\usepackage{epstopdf}
\usepackage{cite}
\usepackage{xcolor}
\begin{document}
%\begin{flushright}
%\underline{\textbf{\today, Version 2}}
%\end{flushright}

%What we need:
%\begin{itemize}
%\item p.5: Supplimentary Material with figure and table
%\item p.5: Discussion and comparison with previous results
%\item Conclusion
%\end{itemize}

%Please mark your changes or inclusions in the text by \textcolor{red}{RED} or \textcolor{blue}{BLUE} colour.

%\clearpage

\title[NO$_2$/WSe$_2$: DFT and PES]{Adsorption of NO$_2$ on WSe$_2$: DFT and photoelectron spectroscopy studies}

\author{R. Ovcharenko$^1$, Yu. Dedkov$^{2,}\footnote{Present address: IHP, Im Technologiepark 25, 15236 Frankfurt (Oder), Germany}$, E. Voloshina$^1$}

\address{$^1$ Humboldt-Universit\"at zu Berlin, Institut f\"ur Chemie, 10099 Berlin, Germany}
\address{$^2$ SPECS Surface Nano Analysis GmbH, Voltastra\ss e 5, 13355 Berlin, Germany}
\ead{elena.voloshina@hu-berlin.de}
\ead{dedkov@ihp-microelectronics.com}

\vspace{10pt}

\begin{abstract}
The electronic structure modifications of WSe$_2$ upon NO$_2$-adsorption at room and low temperatures were studied by means of photoelectron spectroscopy. We found only moderate changes in the electronic structure, which are manifested as an upward shift of the WSe$_2$-related bands to the smaller binding energies. The observed effects are modelled within the density functional theory approach, where a weak adsorption energy of gas molecules on the surface of WSe$_2$ was deduced. The obtained experimental data are explained as a valence bands polarisation effect, which causes their energy shift depending on the adsorption geometry and the formed dipole moment.     
\end{abstract}

%\pacs{73.20.-r, 73.22.Pr, 74.55.+v, 79.60.-i}

\vspace{2pc}
%\noindent{\it Keywords}: TMD, adsorption, DFT, XPS, ARPES

%\submitto{\JPCM}

\maketitle

\ioptwocol

\section{Introduction}

Two dimensional materials, like graphene, $h$-BN, transition metal dichalcogenides (TMDs), etc., which have highest surface-to-volume ratio attract a lot of attention in the fundamental research as well as from the point of the possible technological applications~\cite{Geim:2007a,CastroNeto:2009,Geim:2009,Radisavljevic:2011kx,Wang:2012fa}. For example, for graphene, the most promising ways to use its unique electronic structure (linear dispersion, zero band gap, and zero density of states at the Dirac point, $E_D$) are utilization in flexible touch screens~\cite{Bae:2010,Ryu:2014fo} or gas sensors~\cite{Schedin:2007}. In the later work of Schedin \textit{et al.} it was shown that the $\mu$m-sized graphene-based field-effect transistor shows unprecedented gas sensitivity even at room temperature. This functionality is explained by the change of the resistivity of the graphene sheet upon adsorption or desorption of molecules, which behave as donors or acceptors. The low density of states (DOS) in graphene around $E_D$, which in the neutral state coincides with the Fermi level ($E_F$), allows easily tune the graphene conductivity upon very small charge transfer. Further, it was found that the sensitivity of such graphene-based gas sensors can be improved if dopants or defects are introduced in the graphene matrix~\cite{Zhang:2010,Kumar:2013il,Jiang:2015ia,Lee:2015dx}.

\begin{figure*}[t]
\includegraphics[width=\linewidth]{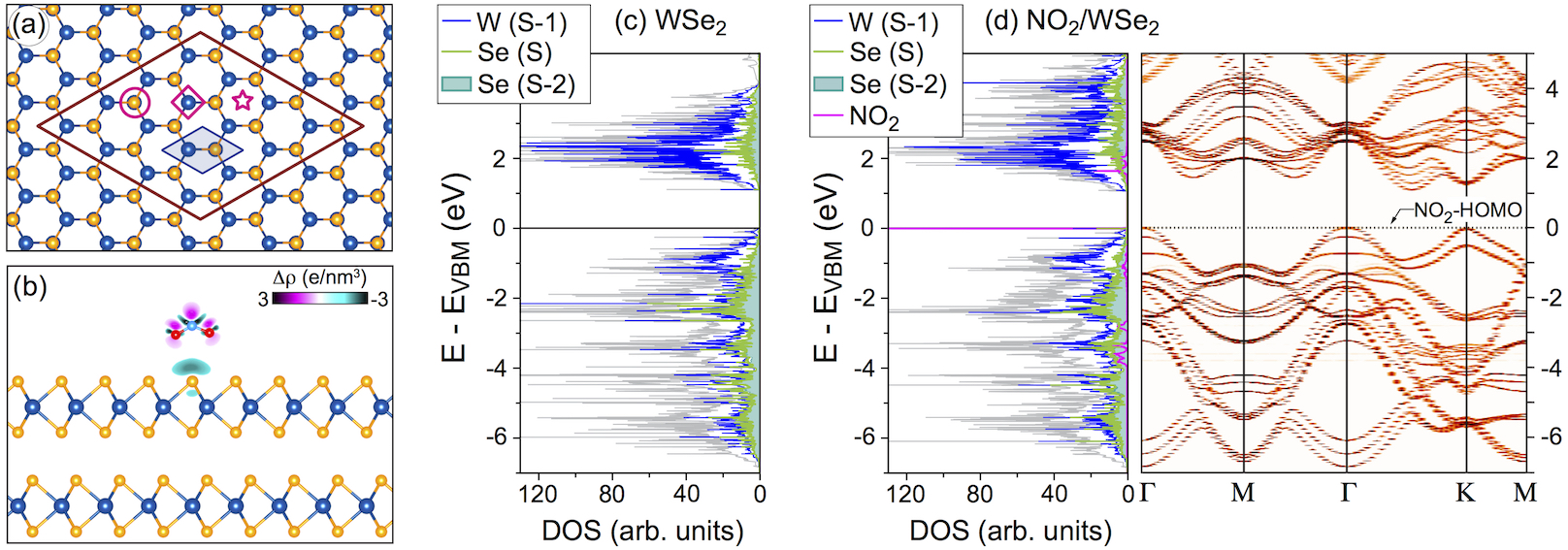}
\caption{(a) Top view of $\textrm{WSe}_2$. The corresponding ($1\times 1$) and ($4\times 4$) unit cells as well as the considered adsopriton positions are marked in the figure. (b) Side view of the energetically most favourable $\textrm{NO}_2/\textrm{WSe}_2$ structure is overlaid with the calculated difference electron density, $\Delta\rho=\rho_{\tiny{\textrm{NO}_2/\textrm{WSe}_2}}-(\rho_{\tiny{\textrm{WSe}_2}}+\rho_{\tiny{\textrm{NO}_2}})$. (c) Total (grey lines) and site-projected density of states calculated for the clean $\textrm{WSe}_2$. (d) Total (grey lines) and site-projected density of states and band structure calculated for the $\textrm{NO}_2/\textrm{WSe}_2$ system in its energetically most favourable adsorprton structure shown in (b).}
\label{DOSandBANDS}
\end{figure*}

Further search for the two-dimensional materials with tuneable transport properties leads to the experiments on the semiconducting transition metal dichalcogenides, with structural formula MX$_2$, where M is a metal atom (Mo or W) and X is a chalcogen (S or Se). These materials have a layered structure, where separated layers are bonded by weak dispersive forces. Each MX$_2$-layer consists of a metal atoms sandwiched between inert layers of chalcogen atoms as shown in Fig.~\ref{DOSandBANDS}(a,b). Covalent bonds localized between metal and chalcogen atoms lead to the semiconducting nature of these materials and to the inert nature of their surfaces at ambient conditions. However, it was shown that they are quite sensitive to the gaseous and higher temperature treatment. For example, MoS$_2$ can be oxidized in the presence of oxygen or/and water already at $100^\circ$C~\cite{Ross:1955ih,KC:2015jy}.  

Recent demonstration of the field-effect transistor (FET) functionality for thin layers of MoS$_2$ and other semiconducting TMDs~\cite{Radisavljevic:2011kx,Wang:2012fa} brings the idea to use these materials in gas sensor devices. The MoS$_2$-based FET with a thickness of $2-4$ single layers used for detection of gases demonstrated high sensitivity for NO gas with a detection limit down to $0.8$\,ppm~\cite{Li:2011il}. Here NO molecules behave as acceptors attracting the electron density from the MoS$_2$ layer. Theoretical analysis of these data confirms the main findings~\cite{Yue:2013km}. In the same work the adsorption of other gases was also considered within the LSDA and GGA approaches and it was found that H$_2$, O$_2$, H$_2$O, NO, and NO$_2$ adsorb weakly on MoS$_2$ and behave like acceptors, whereas NH$_3$ is a charge donor. These theoretical findings were later supported by the recent transport and gas sensing experiments~\cite{Cho:2015ew}.     

Further developments in the filed of electronics and optoelectronics of TMDs bring the W-based materials, like WS$_2$ and WSe$_2$, in the forefront of these studies~\cite{Bhimanapati:2015bo,Gmitra:2015kg,Withers:2015bl}. They have several advantages compared to MoS$_2$, since they have higher thermal stability as well as their electronic structure opens wide perspectives for applications in opto- and spin-electronics, and in the possible new field, the so-called valleytronics, where additional degree of freedom of the electron (valley polarization) might be used for the device operation. The gas sensing properties of WS$_2$ were experimentally studied in Ref.~\cite{Huo:2014cm}, where the influence of the gas adsorption (NH$_3$ and O$_2$) on the photoelectrical properties of the transport devices was investigated at room temperature. Similar to the gas adsorption on MoS$_2$, it was found that molecules are physisorbed on the surface of WS$_2$ with oxygen behaving as acceptor and ammonia leading to the $n$-doping of WS$_2$. With respect to the photoelectrical properties of WS$_2$, $p$- and $n$-doping of this material via molecules adsorption leads to reduction and to increase of photo-responsivity, respectively. Density-functional theory (DFT) analysis (without inclusion of the spin-orbit interaction) of the gas adsorption on WS$_2$~\cite{Zhou:2015hc} shows that the underlying band structure of TMD material is not influenced significantly upon molecules deposition confirming the weak interaction at the interface. In this work it was also found that adsorption of O$_2$, NO, and NO$_2$ on WSe$_2$ leads to the pinning of the lowest unoccupied molecular orbital around the Fermi level. It is interesting that no electronic structure studies by means of photoelectron spectroscopy (of the core levels or valence band states) of gas adsorption on TMDs were published in the literature, motivating the present studies.

Here we present X-ray and angle-resolved photoelectron spectroscopy (XPS and ARPES) results obtained during adsorption of NO$_2$ on WSe$_2$(001). Small changes in the core-levels positions as well as in the valence band were found upon adsorption of gas molecules at room temperature and at $120$\,K indicating the physisorption nature of interaction at the gas-surface interface for TMD material. It is found that adsorption of NO$_2$ leads to the slight $p$-doping of WSe$_2$ as manifested by the rigid shift of core levels and valence band states of TMD to the smaller binding energies. These results are accompanied by the state-of-the-art DFT results, which give the small adsorption energy of NO$_2$ on WSe$_2$ of $E_{ads}=229$\,meV/molecule. The theoretically obtained band structures confirm our experimental findings. 

\section{Methods}

\textit{Theory.} 
The DFT calculations were carried out using the projector augmented wave (PAW) method~\cite{Blochl:1994}, a plane wave basis set and the generalized gradient approximation as parameterized by Perdew \textit{et al.}~\cite{Perdew:1996}, as implemented in the VASP program~\cite{Kresse:1994}. The plane wave kinetic energy cutoff was set to $500$\,eV. The spin-orbit correction, necessary for the proper description of the properties of $\textrm{WSe}_2$, is taken into account via non-collinear magnetism as it is implemented in VASP. The long-range van der Waals (vdW) interactions were accounted for by means of the DFT-D2 approach~\cite{Grimme:2006}. The supercell used in this work has a ($4\times4$) lateral periodicity with respect to $\textrm{WSe}_2$. It is constructed from a slab of 3 layers of $\textrm{WSe}_2$ with a single $\textrm{NO}_2$ molecule adsorbed from one side and a vacuum region of approximately $20$\,\AA. In the total energy calculations and during the structural relaxations the $k$-meshes for sampling the supercell Brillouin zone are chosen to be as dense as $3\times3$ and $6\times6$, respectively, and centred at the $\Gamma$-point. The band structures calculated for the studied systems were unfolded to the $\textrm{WSe}_2$ ($1\times1$) primitive unit cell according to the procedure described in Refs.~\cite{Popescu:2012bq,Medeiros:2014ka} with the code BandUP~\cite{Medeiros:2014ka}.

\begin{figure}[b]
\includegraphics[width=\linewidth]{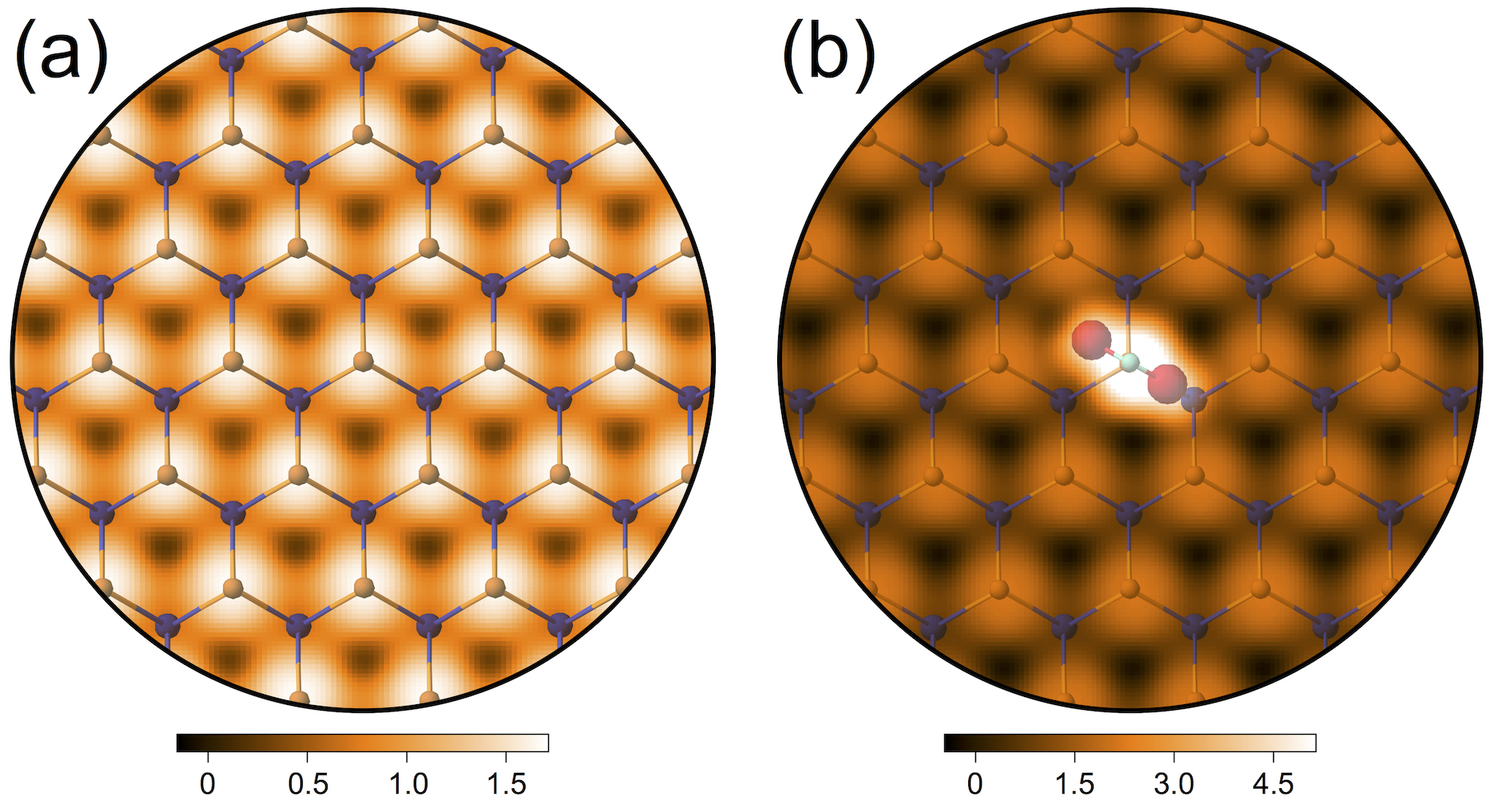}
\caption{Calculated STM images of (a) WSe$_2$(0001) and (b) NO$_2$/WSe$_2$(0001). Charge integration was performed in the energy range of $1$\,eV below the middle of the energy gap of WSe$_2$ (see Fig.~\ref{DOSandBANDS}). Corresponding height color scales (in \AA) are shown in the lower parts.}
\label{STM}
\end{figure}

\textit{Experiment.} The XPS and ARPES measurements  with Al\, $K\alpha$ ($h\nu=1486.6$\,eV) and He\,I$\alpha$ radiation ($h\nu=21.2$\,eV), respectively, were performed using a photoemission system with PHOIBOS 150 analyzer equipped with 2D-CCD detector. In this case a 5-axis motorized manipulator was used, allowing for a precise alignment of the sample in the $k$ space for the ARPES experiments. The sample was azimuthally pre-aligned in such a way that the polar scans were performed along the $\Gamma-\mathrm{K}$ or $\Gamma-\mathrm{M}$ directions of the hexagonal Brillouine zone of WSe$_2$ with the photoemission intensity on the channelplate images acquired along the direction perpendicular to the respective scanning direction. The final 3D data sets of the photoemission intensity as a function of kinetic energy and two emission angles, $I(E_{kin},angle_1,angle_2)$, were then carefully analyzed. The base vacuum in the experimental station is in the range of $1\times10^{-10}$\,mbar.

WSe$_2$(0001) samples (size of approximately $2\times2\times0.2$\,mm$^3$) were pressed to the Mo sample holder with Ta clamps, which were used for the careful Fermi level calibration. Samples were cleaved in air and then were introduced in vacuum within the next $30$\,sec. Prior every series of XPS or ARPES experiments WSe$_2$ samples were carefully degassed at $300^\circ$\,C for $30$\,min. Dosing of NO$_2$ (measured in langmuir (L), $1\mathrm{L}=1\cdot10^{-6}\,\mathrm{Torr}\times1\,\mathrm{sec}$) was performed via electronically controlled leak-valve and sample was kept either at room temperature or at $120$\,K.       

\section{Results and discussions}

Adsorption of NO$_2$ molecules on the WSe$_2$(0001) surface was studied within the DFT approach using PBE functional and dispersive forces (van der Waals) were taken into account via formalism proposed by Grimme~\cite{Grimme:2006}. Since the studied TMD material contains heavy elements, then the spin-orbit interaction was included in our calculations. Fig.~\ref{DOSandBANDS}(a,b) shows the top and side views of the crystallographic structure of WSe$_2$, where high symmetry adsorption places for NO$_2$ are marked by the respective symbols. Our calculations for the NO$_2$/WSe$_2$ system show that for all considered configurations the adsorption energy is in the range of $148-229$\,meV per NO$_2$-molecule (see Fig.\,S1 and Table\,T1 in the Supplementary material where all configurations with adsorption energies are listed) placing this system to the class of weakly-bonded one, where adsorption is governed predominantly by the dispersion interaction. In our further analysis we will consider only one configuration, namely when NO$_2$ molecule is paced above Se (in the ($4\times4$) configuration resulting $0.06$\,ML coverage) where N-atom is located directly above Se-atom and both N-O bonds are directed downwards with one of them along the Se-W bond and other pointed to the centre of the W-Se ``hexagon''. Taking into account the physisorption nature of the bonding in this system we can expect the similar results for other geometrical arrangement of NO$_2$ molecules on WSe$_2$. 

Figure~\ref{DOSandBANDS}(c) shows calculated total and atom-projected partial density of states (DOS and PDOS) of the studied systems as well as band structure obtained for the energetically most favourable $\textrm{NO}_2/\textrm{WSe}_2$ structure. The calculated electronic structure for WSe$_2$ is in very good agreement with previously published theoretical and experimental data~\cite{Finteis:1997vs,Jiang:2012kv,Yuan:2013bc,Riley:2014bw,Le:2015hj} as well as with our ARPES results presented later. Our calculations show that bulk WSe$_2$ is an indirect semiconducting material with a band gap of $1.1$\,eV. The obtained spin-orbit splitting at the $K$-point of the Brillouine zone is $485$\,meV.

\begin{figure}%[t]
\includegraphics[width=\linewidth]{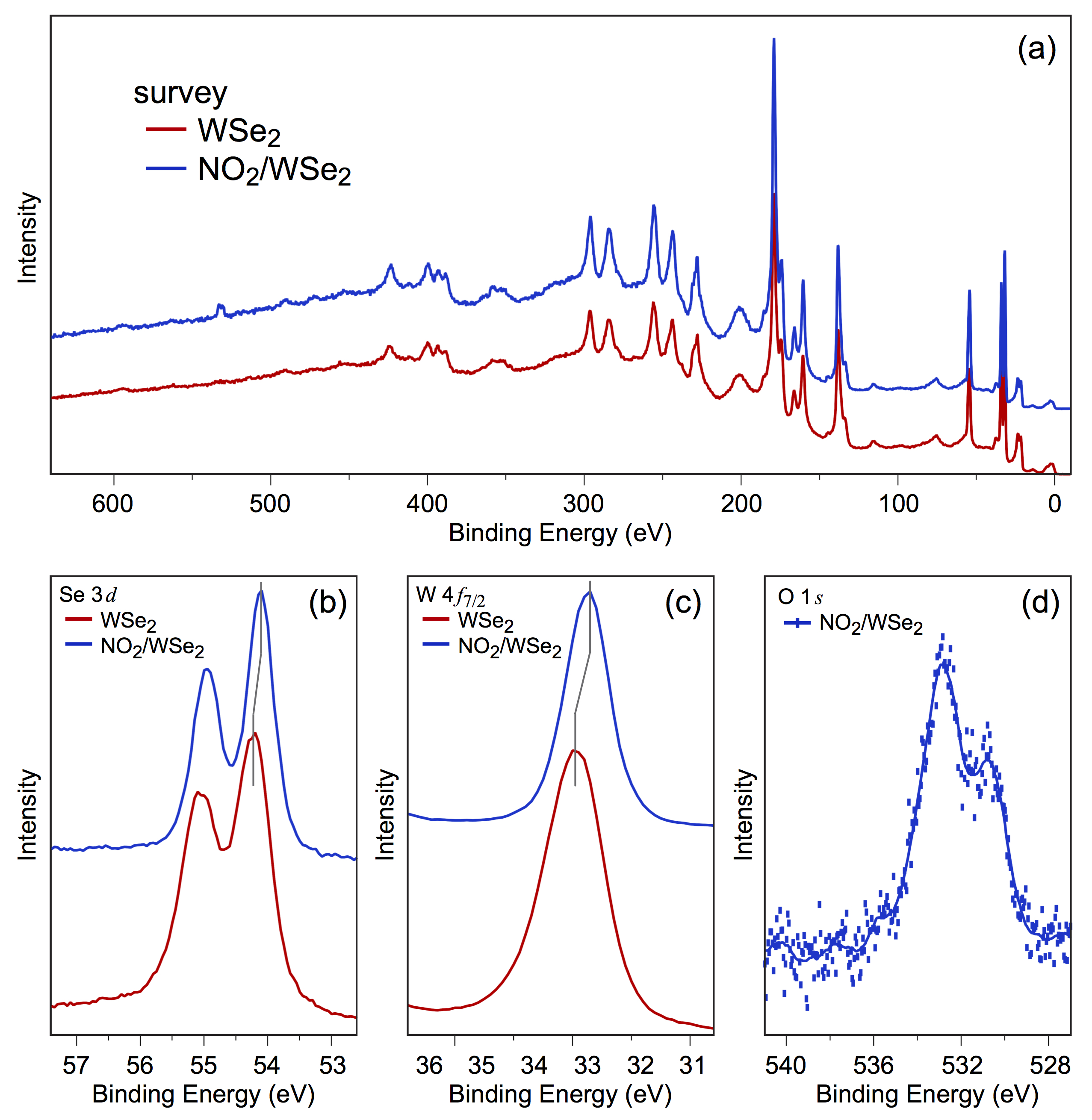}
\caption{XPS spectra collected before (brown) and after (blue) adsorption of $600$\,L of NO$_2$ on WSe$_2$: (a) survey, (b) Se\,$3d$, (c) W\,$4f_{7/2}$, and (d) O\,$1s$.}
\label{NO2onWSe2_XPS}
\end{figure}

Adsorption of NO$_2$ on the surface of WSe$_2$ leads to the formation of the impurity molecule-induced band in the vicinity of the top of the valence band of WSe$_2$. Other NO$_2$-related bands are found in the energy ranges at $E-E_F\approx-3\dots-4$\,eV and $E-E_F\approx1.5\dots2$\,eV. However, as one can see from the presented DOS and band structure of the NO$_2$/WSe$_2$ system (Fig.~\ref{DOSandBANDS}(d)) there is no \textit{hybridization} between valence band states of WSe$_2$-substrate and molecular orbitals of NO$_2$. Adsorption of NO$_2$ leads only to the charge transfer from substrate on the molecule as can be seen in Fig.~\ref{DOSandBANDS}(b), where charge depletion (accumulation) is found on Se-atoms (NO$_2$ molecules). Such effect leads to the polarization of the electronic states of substrate that can be detected in the photoemission experiments.

Weak adsorption of NO$_2$ on WSe$_2$(0001) is also revealed in the calculated STM images (Fig.~\ref{STM}). In these calculations, performed in the framework of the Tersoff-Hamann formalism~\cite{Tersoff:1985}, the charge density integration was performed for the energy range of $1$\,eV below of the middle of the energy gap of WSe$_2$. In this case the NO$_2$ HOMO state is included in the integration energy range for the NO$_2$/WSe$_2$ system. One can see that adsorption of NO$_2$ on the surface of the TMD material does not change the local density of states that can be taken as a signature of the absence of the orbital hybridization of the molecular orbitals and valence band states of WSe$_2$.

The behaviour of isolated NO$_2$ molecule on WSe$_2$ is similar to the one on the other TMD surfaces like MoS$_2$ or WS$_2$. Indeed, in all cases the molecule-surface interaction is accounted for by weak dispersion forces, which do not drastically affect the electronic structure of adsorbent and result in relatively small adsorption energies: $229$\,meV, $412$\,meV~\cite{Zhou:2015hc}, and $140$\,meV\cite{Cho:2015ew} for WSe$_2$, WS$_2$ and MoS$_2$, respectively. The adsorption geometry where oxygens of NO$_2$ molecule are directed towards surface was found to be the most energetically favourable also for adsorption on the top of all these  surfaces. However, the adsorption positions were found to be different: while for WS$_2$ and MoS$_2$ the molecule tends to adsorb above center of the ring, in WSe$_2$ the site right above surface Se atom was found to be the most favourable. It should be noted here, the deviation in adsorption energies at different sites in the latter case is just about few meV that is, in principle, at the limit of DFT calculation precision. The important point is that in all cases NO$_2$ acts as electron acceptor that helps to change the conductivity of surface upon NO$_2$ adsorption in wide limits and opens up the potential opportunities to employ TMD materials as sensitive NO$_2$ gas detector. During adsorption the surface valence electron occupies the LUMO of NO$_2$ which finally turns out to be located around Fermi level of combined NO$_2$/{\rm TMD} system. Such effect of the localisation of the flat adsorbate-induced impurity state at the Fermi level of TMD surface was early observed for a number of adsorbates like O$_2$, NO, NO$_2$ on different TMD materials and was called Fermi-level pinning phenomenon~\cite{Zhou:2015hc}. As was shown in latter work such effect may be explained within traditional charge transfer theory provided isolated molecular LUMO is lower in energy than Fermi level of pristine surface. That causes the drift of electron density from the surface to unoccupied well-localised molecular orbital without hybridisation or formation of a strong chemical bonding between adsorbate and adsorbent allowing only weak dispersion electrostatic attraction.

\begin{figure}%[t]
\includegraphics[width=\linewidth]{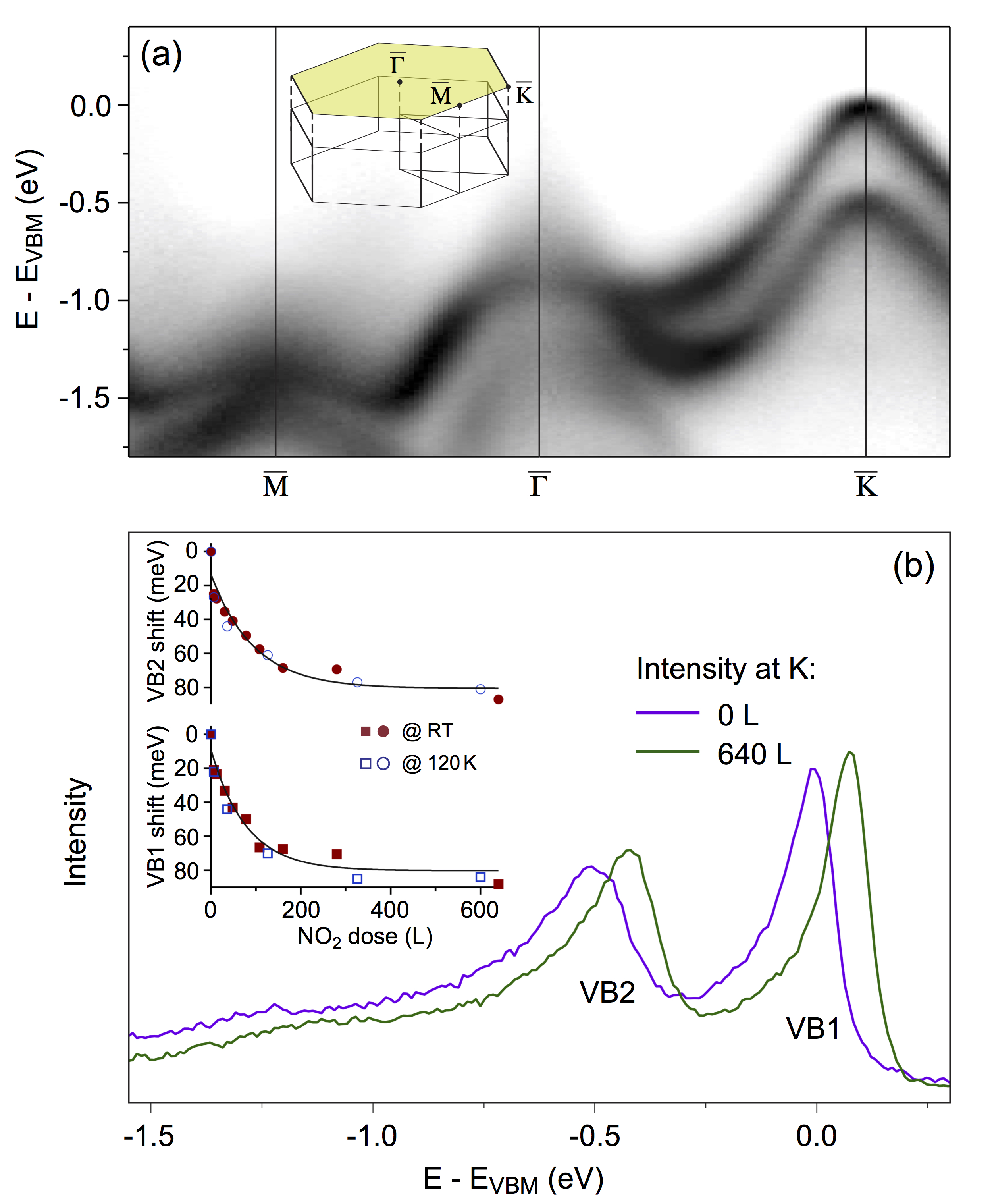}
\caption{(a) Band structure of WSe$_2$ along the high-symmetry directions of the hexagonal Brillouine zone (shown as an inset) extracted from the complete 3D data set for the ARPES intensity. (b) Intensity profiles extracted at the $\overline{\mathrm{K}}$-point for the data collected before (pink) and after (green) adsorption of $640$\,L of NO$_2$ on WSe$_2$. Insets show the respective energy shift of VB1 and VB2 as a function of NO$_2$ dose. Two data sets are presented, which were collected at room temperature and $120$\,K.}
\label{ARPES_Doping}
\end{figure}

In order to proof the theoretical findings on the weak adsorption of NO$_2$ on WSe$_2$ we perform systematic XPS and APRES experiments for this system. Experiments were performed with the WSe$_2$ substrate kept either at $120$\,K or $300$\,K (RT). Fig.~\ref{NO2onWSe2_XPS} shows (a) survey, (b) Se\,$3d$, (c) W\,$4f_{7/2}$ XPS spectra of WSe$_2$ before and after dosing of $600$\,L of NO$_2$ at $120$\,K. The O\,$1s$ XPS spectrum after dosing is shown in panel (d) (the N\,$1s$ emission line overlaps in energy with Se Auger lines). One can clearly see that adsorption of NO$_2$ on WSe$_2$ leads to the minimal modifications of the TMD-related emission lines. The energy shift of W\,$4f$ and Se\,$3d$ lines of $100\pm10$\,meV to the smaller binding energies is detected indicating a weak interaction between adsorbed NO$_2$ and underlying WSe$_2$. There are no additional low or high binding energy components in the respective XPS spectra pointing the absence of the strong chemical interaction in the system. As was discussed in the previous paragraph and can be concluded from the presented XPS data the interaction in the present adsorption system has vdW character that leads only to the polarization of the upper WSe$_2$ layer due to the dipole moment of the adsorbed NO$_2$ molecules. Such interaction is manifested as a rigid shift of all substrate-related emission lines to the lower binding energies.

The discussed effect of polarization of the top layer of TMD material was also detected in our ARPES experiments on WSe$_2$ upon adsorption of NO$_2$ molecules at RT and $120$\,K. Fig.~\ref{ARPES_Doping}(a) shows the ARPES photoemission map for clean WSe$_2$ along the $\overline{\mathrm{M}}-\overline{\Gamma}-\overline{\mathrm{K}}$ directions of the hexagonal Brillouine zone (shown as an inset) extracted from the complete 3D data set acquired at $120$\,K. These data are in very good agreement with the band structure calculations presented earlier [Fig.~\ref{DOSandBANDS}(a)] and with previous data~\cite{Riley:2014bw,Le:2015hj}. The extracted value of spin-orbit splitting between two bands (marked as VB1 and VB2) at the $\overline{\mathrm{K}}$ point is $503$\,meV giving a very good agreement with the respective value obtained in the calculated band structure of WSe$_2$.

Adsorption of NO$_2$ on WSe$_2$(0001) surface leads to monotonic energy shift of the bands at the $\overline{\mathrm{K}}$ point as a function of the NO$_2$ and reaches the maximal value of $87\pm5$\,meV at $640$\,L of the NO$_2$ dose [Fig.~\ref{DOSandBANDS}(b)]. (For every NO$_2$ dose the Fermi level position was carefully calibrated via measurement of the PES spectra of the Ta clamps allowing to trace the shift of the valence band states of WSe$_2$.) Here we would like to emphasise that the measured energy shift does not depend on the temperature of the WSe$_2$ substrate as can be seen from the presented dependencies of the energy positions of VB1 and VB2 bands as a function of the gas dose [inset of Fig.~\ref{DOSandBANDS}(b)]. Taking into account that energy shifts for the core-level lines and valence band states, measured by XPS and ARPES, respectively, have similar values we can conclude here on the weak physisorption of NO$_2$ molecules on WSe$_2$ leading to the polarization of the top layer of TMD substrate due to the dipole moment of NO$_2$ molecules. We can see that the studied surface of WSe$_2$ is extremely inert preventing further oxidation of the bulk of TMD material even at very high dose of the strong oxidation gas. This effect can be explained by the formation of the strong covalent-like bonds between W and Se in the layer, leading to the charge accumulation inside the layer and preventing any strong interactions on the surface. The presented experimental results are also supported by our DFT calculations which conclude on the weak interaction between NO$_2$ and WSe$_2$.

\section{Conclusions}

We present here the comprehensive study of the NO$_2$ adsorption on the WSe$_2$ surface by means of a number experimental spectroscopic techniques like XPS and ARPES supported by accurate spin-polarised DFT-D2 electronic structure calculations. At each step our theoretical and experimental results are in well agreement with each other that indicates in reasonable theoretical model and experimental conditions. Experimentally, the weak physisorption of NO$_2$ molecules on WSe$_2$ leading to the polarization of the top layer of TMD substrate due to the dipole moment of NO$_2$ molecules has been found. These findings were then confirmed by means of DFT calculation showing us the absence of any hybridisation between NO$_2$ molecular orbitals with WSe$_2$ bands and filling of the NO$_2$ LUMO with surface electron density. The latter effect, which is also closely related to the so-called Fermi-level pinning phenomenon, may be easily explained within traditional charge transfer theory and has been early observed for adsorption of the O$_2$, NO and NO$_2$ molecules on the top of different TMD surfaces.
 
\section*{Acknowledgements}

The High Performance Computing Network of Northern Germany (HLRN-III) is acknowledged for computer time. Financial support from the German Research Foundation (DFG) through the grant VO1711/3-1 within the Priority Programme 1459 ``Graphene'' is appreciated.

\section*{References}

%\bibliography{/Users/YuDedkov/Work/Articles/___REFERENCES___/references_all.bib}

\begin{thebibliography}{10}
\expandafter\ifx\csname url\endcsname\relax
  \def\url#1{{\tt #1}}\fi
\expandafter\ifx\csname urlprefix\endcsname\relax\def\urlprefix{URL }\fi
\providecommand{\eprint}[2][]{\url{#2}}

\bibitem{Geim:2007a}
Geim A~K and Novoselov K~S 2007 {\em Nature Mater.\/} {\bf 6} 183--191

\bibitem{CastroNeto:2009}
Castro~Neto A, Guinea F, Peres N, Novoselov K and Geim A 2009 {\em Rev. Mod.
  Phys.\/} {\bf 81} 109--162

\bibitem{Geim:2009}
Geim A 2009 {\em Science\/} {\bf 324} 1530--1534

\bibitem{Radisavljevic:2011kx}
Radisavljevic B, Radenovic A, Brivio J, Giacometti V and Kis A 2011 {\em Nature
  Nanotech.\/} {\bf 6} 147

\bibitem{Wang:2012fa}
Wang Q~H, Kalantar-zadeh K, Kis A, Coleman J~N and Strano M~S 2012 {\em Nature
  Nanotech.\/} {\bf 7} 699--712

\bibitem{Bae:2010}
Bae S, Kim H, Lee Y, Xu X, Park J~S, Zheng Y, Balakrishnan J, Lei T, Kim H~R,
  Song Y~I, Kim Y~J, Kim K~S, Ozyilmaz B, Ahn J~H, Hong B~H and Iijima S 2010
  {\em Nature Nanotech.\/} {\bf 5} 574--578

\bibitem{Ryu:2014fo}
Ryu J, Kim Y, Won D, Kim N, Park J~S, Lee E~K, Cho D, Cho S~P, Kim S~J, Ryu
  G~H, Shin H~A~S, Lee Z, Hong B~H and Cho S 2014 {\em ACS Nano\/} {\bf 8}
  950--956

\bibitem{Schedin:2007}
Schedin F, Geim A~K, Morozov S~V, Hill E~W, Blake P, Katsnelson M~I and
  Novoselov K~S 2007 {\em Nature Mater.\/} {\bf 6} 652--655

\bibitem{Zhang:2010}
Zhang Y~H, Zhou K~G, Xie K~F, Zeng J, Zhang H~L and Peng Y 2010 {\em
  Nanotechnology\/} {\bf 21} 065201

\bibitem{Kumar:2013il}
Kumar B, Min K, Bashirzadeh M, Farimani A~B, Bae M~H, Estrada D, Kim Y~D,
  Yasaei P, Park Y~D, Pop E, Aluru N~R and Salehi-Khojin A 2013 {\em Nano
  Lett.\/} {\bf 13} 1962--1968

\bibitem{Jiang:2015ia}
Jiang Y, Yang S, Li S, Liu W and Zhao Y 2015 {\em J. Nanomater.\/}
  {\bf 2015} 504103

\bibitem{Lee:2015dx}
Lee G, Yang G, Cho A, Han J~W and Kim J 2015 {\em Phys. Chem. Chem. Phys.\/}; doi: 10.1039/C5CP04422G

\bibitem{Ross:1955ih}
Ross S and Sussman A 1955 {\em J. Phys. Chem.\/} {\bf 59}
  889--892

\bibitem{KC:2015jy}
KC S, Longo R~C, Wallace R~M and Cho K 2015 {\em J. Appl. Phys.\/} {\bf 117}
  135301

\bibitem{Li:2011il}
Li H, Yin Z, He Q, Li H, Huang X, Lu G, Fam D~W~H, Tok A~I~Y, Zhang Q and Zhang
  H 2011 {\em Small\/} {\bf 8} 63--67

\bibitem{Yue:2013km}
Yue Q, Shao Z, Chang S and Li J 2013 {\em Nanoscale Res. Lett.\/} {\bf 8}
  425

\bibitem{Cho:2015ew}
Cho B, Hahm M~G, Choi M, Yoon J, Kim A~R, Lee Y~J, Park S~G, Kwon J~D, Kim C~S,
  Song M, Jeong Y, Nam K~S, Lee S, Yoo T~J, Kang C~G, Lee B~H, Ko H~C, Ajayan
  P~M and Kim D~H 2015 {\em Sci. Rep.\/} {\bf 5} 8052

\bibitem{Bhimanapati:2015bo}
Bhimanapati G~R, Lin Z, Meunier V, Jung Y, Cha J, Das S, Xiao D, Son Y, Strano
  M~S, Cooper V~R, Liang L, Louie S~G, Ringe E, Zhou W, Kim S~S, Naik R~R,
  Sumpter B~G, Terrones H, Xia F, Wang Y, Zhu J, Akinwande D, Alem N, Schuller
  J~A, Schaak R~E, Terrones M and Robinson J~A 2015 {\em ACS Nano\/} {\bf 9}
  11509--11539

\bibitem{Gmitra:2015kg}
Gmitra M and Fabian J 2015 {\em Phys. Rev. B\/} {\bf 92} 155403--6

\bibitem{Withers:2015bl}
Withers F, Del Pozo-Zamudio O, Mishchenko A, Rooney A~P, Gholinia A, Watanabe
  K, Taniguchi T, Haigh S~J, Geim A~K, Tartakovskii A~I and Novoselov K~S 2015
  {\em Nature Mater.\/} {\bf 14} 301--306

\bibitem{Huo:2014cm}
Huo N, Yang S, Wei Z, Li S~S, Xia J~B and Li J 2014 {\em Sci. Rep.\/} {\bf 4}
  5209

\bibitem{Zhou:2015hc}
Zhou C, Yang W and Zhu H 2015 {\em J. Chem. Phys.\/} {\bf 142}
  214704--9

\bibitem{Blochl:1994}
Bl{\"o}chl P~E 1994 {\em Phys. Rev. B\/} {\bf 50} 17953--17979

\bibitem{Perdew:1996}
Perdew J, Burke K and Ernzerhof M 1996 {\em Phys. Rev. Lett.\/} {\bf 77}
  3865--3868

\bibitem{Kresse:1994}
Kresse G and Hafner J 1994 {\em J. Phys.: Condens. Matter\/} {\bf 6} 8245--8257

\bibitem{Grimme:2006}
Grimme S 2006 {\em J. Comput. Chem.\/} {\bf 27} 1787--1799

\bibitem{Popescu:2012bq}
Popescu V and Zunger A 2012 {\em Phys. Rev. B\/} {\bf 85} 085201

\bibitem{Medeiros:2014ka}
Medeiros P~V~C, Stafstr{\"o}m S and Bj{\"o}rk J 2014 {\em Phys. Rev. B\/} {\bf
  89} 041407

\bibitem{Finteis:1997vs}
Finteis T, Hengsberger M, Straub T, Fauth K, Claessen R, Auer P, Steiner P,
  Hufner S, Blaha P and V{\"o}gt M 1997 {\em Phys. Rev. B\/} {\bf 55} 10400

\bibitem{Jiang:2012kv}
Jiang H 2012 {\em J. Phys. Chem. C\/} {\bf 116} 7664--7671

\bibitem{Yuan:2013bc}
Yuan H 2013 {\em Nature Phys.\/} {\bf 9} 563--569

\bibitem{Riley:2014bw}
Riley J~M, Mazzola F, Dendzik M, Michiardi M, Takayama T, Bawden L, Graner{\o}d
  C, Leandersson M, Balasubramanian T, Hoesch M, Kim T~K, Takagi H, Meevasana
  W, Hofmann P, Bahramy M~S, Wells J~W and King P~D~C 2014 {\em Nature
  Phys.\/} {\bf 10} 835--839

\bibitem{Le:2015hj}
Le D, Barinov A, Preciado E, Isarraraz M, Tanabe I, Komesu T, Troha C, Bartels
  L, Rahman T~S and Dowben P~A 2015 {\em J. Phys.: Condens. Matter\/}  182201

\bibitem{Tersoff:1985}
Tersoff J and Hamann D~R 1985 {\em Phys. Rev. B\/} {\bf 31} 805--813

\end{thebibliography}

\providecommand{\newblock}{}

\clearpage

%\begin{flushright}
%prepared for \textit{Appl. Phys. Lett.}\\
%Version\#5
%size is restricted to 3 (!!!) two-columns pp.;\\
%will be shortened later
%\end{flushright}
\noindent
Supplementary material for the manuscript:\\

\noindent\textbf{
Adsorption of NO$_2$ on WSe$_2$: DFT and photoelectron spectroscopy studies
}\\
\newline
R. Ovcharenko$^1$, Yu. Dedkov$^2$, and E. Voloshina$^1$
\\
\newline
\small
$^1$ Humboldt Universit\"at zu Berlin, Institut f\"ur Chemie, 12489 Berlin, Germany\\
$^2$SPECS Surface Nano Analysis GmbH, Voltastra\ss e 5, 13355 Berlin, Germany
\newline
E-mail: elena.voloshina@hu-berlin.de\\
E-mail: dedkov@ihp-microelectronics.com
\vspace{1cm}
\newline
\normalsize
\noindent \textbf{Content:}
\begin{itemize}

\item[1.]  Table T1: Calculated adsorption energies of the NO$_2$/WSe$_2$ system.

\item[2.]  Figure S1: Considered adsorption geometries of the NO$_2$/WSe$_2$ system.  

\end{itemize}

\linespread{1.2}

\vspace{1cm}

\clearpage
\noindent\textbf{Table\,S1.} Calculated adsorption energies (in meV) of the NO$_2$ molecule on the WSe$_2$ surface for geometries shown in Figure S1. 
\begin{center}
\begin{tabular}{ |c|c|c|c| } 
\hline
 & \textbf{\qquad geom-1 \qquad} & \textbf{\qquad geom-2 \qquad} & \textbf{\qquad geom-3 \qquad} \\ 
\hline
\qquad\textbf{Ring\qquad} & -160 & -223 & -153 \\ 
\hline
\textbf{Se} & -148 & -229 & -147 \\ 
\hline
\textbf{W} & -155 & -224 & -149 \\
\hline
\end{tabular}
\end{center}

\clearpage
\begin{figure*}[t]
\includegraphics[width=0.75\linewidth]{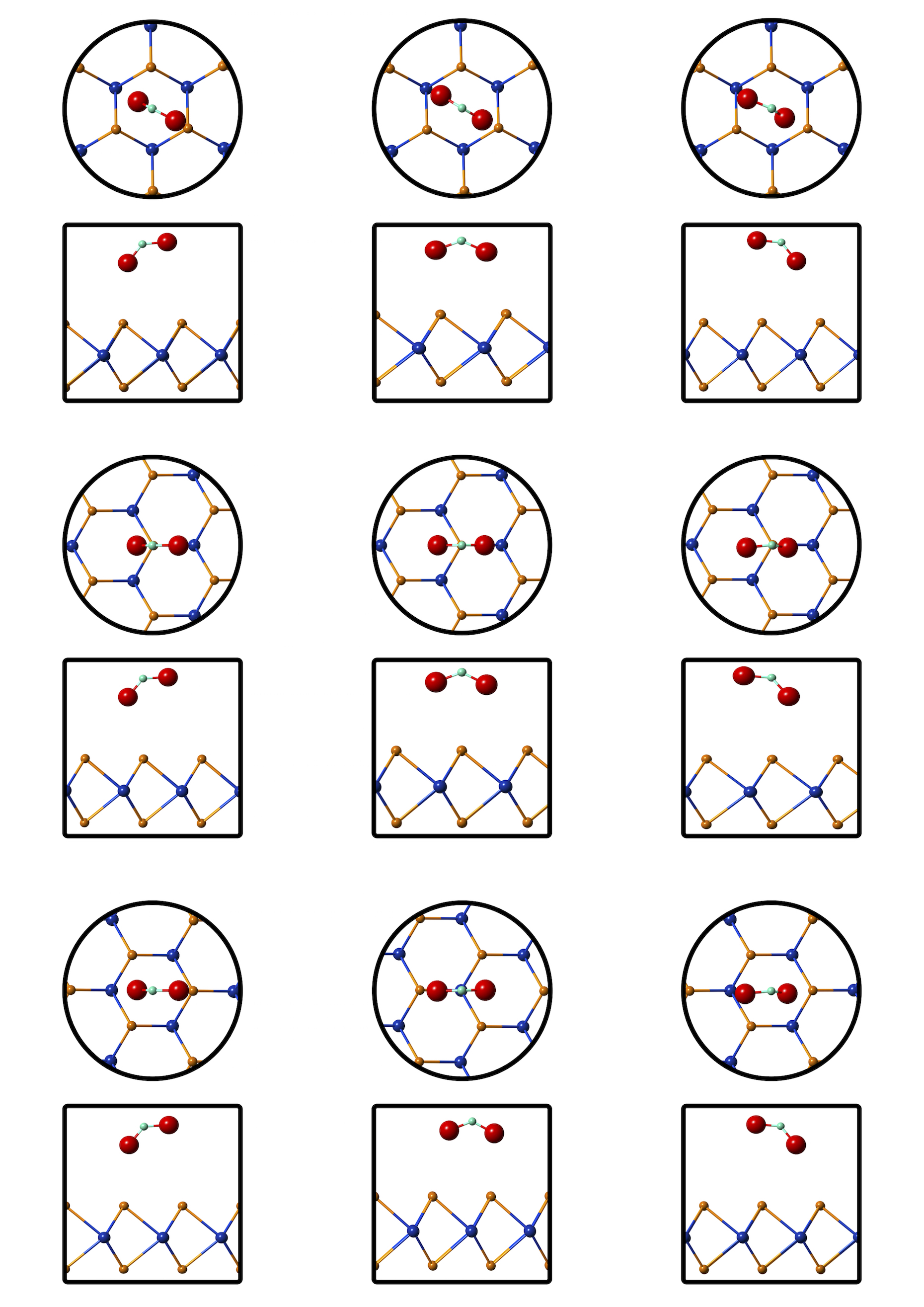}
\caption{\textbf{Figure\,S1.} Adsorption geometries of the NO$_2$/WSe$_2$ system after ionic relaxation. In total we have considered three adsorption places: above the ring (top), above Se atom (middle) and above W atom (bottom) lines. For each place three molecule orientations were taken into account: one molecule oxygen is directed towards surface and center of the ring (geom-1 -- left), both molecule oxygens are directed towards surface (geom-2 -- middle) and one molecule oxygen is directed towards one of the surface site (geom-3 -- right) columns.}
\end{figure*}

\end{document}